\documentclass[10pt,a4paper,onecolumn]{article}
\usepackage{marginnote}
\usepackage{graphicx}
\usepackage{xcolor}
\usepackage{authblk,etoolbox}
\usepackage{titlesec}
\usepackage{calc}
\usepackage{tikz}
\usepackage{hyperref}
\hypersetup{colorlinks,breaklinks=true,
            urlcolor=[rgb]{0.0, 0.5, 1.0},
            linkcolor=[rgb]{0.0, 0.5, 1.0}}
\usepackage{caption}
\usepackage{tcolorbox}
\usepackage{amssymb,amsmath}
\usepackage{ifxetex,ifluatex}
\usepackage{seqsplit}
\usepackage{xstring}

\usepackage{float}
\let\origfigure\figure
\let\endorigfigure\endfigure
\renewenvironment{figure}[1][2] {
    \expandafter\origfigure\expandafter[H]
} {
    \endorigfigure
}

\usepackage{fixltx2e} 
\usepackage[
  backend=biber,
]{biblatex}
\bibliography{paper.bib}

\let\textttOrig=\texttt
\def\texttt#1{\expandafter\textttOrig{\seqsplit{#1}}}
\renewcommand{\seqinsert}{\ifmmode
  \allowbreak
  \else\penalty6000\hspace{0pt plus 0.02em}\fi}

\makeatletter
\let\href@Orig=\href
\def\href@Urllike#1#2{\href@Orig{#1}{\begingroup
    \def\Url@String{#2}\Url@FormatString
    \endgroup}}
\def\href@Notdoi#1#2{\def\tempa{#1}\def\tempb{#2}%
  \ifx\tempa\tempb\relax\href@Urllike{#1}{#2}\else
  \href@Orig{#1}{#2}\fi}
\def\href#1#2{%
  \IfBeginWith{#1}{https://doi.org}%
  {\href@Urllike{#1}{#2}}{\href@Notdoi{#1}{#2}}}
\makeatother

\newlength{\cslhangindent}
\newlength{\csllabelwidth}
\newenvironment{CSLReferences}[3] 
 {
  \setlength{\parindent}{0pt}
  \ifodd #1 \everypar{\setlength{\hangindent}{\cslhangindent}}\ignorespaces\fi
  \ifnum #2 > 0
  \setlength{\parskip}{#2\baselineskip}
  \fi
 }%
 {}
\usepackage{calc}

\usepackage[top=3.5cm, bottom=3cm, right=1.5cm, left=1.0cm,
            headheight=2.2cm, reversemp, includemp, marginparwidth=4.5cm]{geometry}

\titleformat{\section}
  {\normalfont\sffamily\Large\bfseries}
  {}{0pt}{}
\titleformat{\subsection}
  {\normalfont\sffamily\large\bfseries}
  {}{0pt}{}
\titleformat{\subsubsection}
  {\normalfont\sffamily\bfseries}
  {}{0pt}{}
\titleformat*{\paragraph}
  {\sffamily\normalsize}

\usepackage{fancyhdr}
\makeatletter
\let\ps@plain\ps@fancy
\fancyheadoffset[L]{4.5cm}
\fancyfootoffset[L]{4.5cm}

\definecolor{linky}{rgb}{0.0, 0.5, 1.0}

\newtcolorbox{repobox}
   {colback=red, colframe=red!75!black,
     boxrule=0.5pt, arc=2pt, left=6pt, right=6pt, top=3pt, bottom=3pt}

\newcommand{\ExternalLink}{%
   \tikz[x=1.2ex, y=1.2ex, baseline=-0.05ex]{%
       \begin{scope}[x=1ex, y=1ex]
           \clip (-0.1,-0.1)
               --++ (-0, 1.2)
               --++ (0.6, 0)
               --++ (0, -0.6)
               --++ (0.6, 0)
               --++ (0, -1);
           \path[draw,
               line width = 0.5,
               rounded corners=0.5]
               (0,0) rectangle (1,1);
       \end{scope}
       \path[draw, line width = 0.5] (0.5, 0.5)
           -- (1, 1);
       \path[draw, line width = 0.5] (0.6, 1)
           -- (1, 1) -- (1, 0.6);
       }
   }

\patchcmd{\@maketitle}{center}{flushleft}{}{}
\patchcmd{\@maketitle}{center}{flushleft}{}{}
\patchcmd{\@maketitle}{\LARGE}{\LARGE\sffamily}{}{}
\def\maketitle{{%
  
  \AB@maketitle}}
\makeatletter
\renewcommand\AB@affilsepx{ \protect\Affilfont}
\renewcommand\AB@affilnote[1]{{\bfseries #1}\hspace{3pt}}
\renewcommand{\affil}[2][]%
   {\newaffiltrue\let\AB@blk@and\AB@pand
      \if\relax#1\relax\def\AB@note{\AB@thenote}\else\def\AB@note{#1}%
        \setcounter{Maxaffil}{0}\fi
        \begingroup
        \let\href=\href@Orig
        \let\texttt=\textttOrig
        \let\protect\@unexpandable@protect
        \def\thanks{\protect\thanks}\def\footnote{\protect\footnote}%
        \@temptokena=\expandafter{\AB@authors}%
        {\def\\{\protect\\\protect\Affilfont}\xdef\AB@temp{#2}}%
         \xdef\AB@authors{\the\@temptokena\AB@las\AB@au@str
         \protect\\[\affilsep]\protect\Affilfont\AB@temp}%
         \gdef\AB@las{}\gdef\AB@au@str{}%
        {\def\\{, \ignorespaces}\xdef\AB@temp{#2}}%
        \@temptokena=\expandafter{\AB@affillist}%
        \xdef\AB@affillist{\the\@temptokena \AB@affilsep
          \AB@affilnote{\AB@note}\protect\Affilfont\AB@temp}%
      \endgroup
       \let\AB@affilsep\AB@affilsepx
}
\makeatother

\renewcommand\Affilfont{\sffamily\small\mdseries}
\ifnum 0\ifxetex 1\fi\ifluatex 1\fi=0 
  \usepackage[T1]{fontenc}
  \usepackage[utf8]{inputenc}

\else 
  \ifxetex
    \usepackage{mathspec}
    \usepackage{fontspec}

  \else
    \usepackage{fontspec}
  \fi
  \defaultfontfeatures{Ligatures=TeX,Scale=MatchLowercase}

\fi
\IfFileExists{upquote.sty}{\usepackage{upquote}}{}
\IfFileExists{microtype.sty}{%
\usepackage{microtype}
\UseMicrotypeSet[protrusion]{basicmath} 
}{}

\usepackage{hyperref}
\hypersetup{unicode=true,
            pdftitle={radioactivedecay: A Python package for radioactive decay calculations},
            pdfborder={0 0 0},
            breaklinks=true}
\let\addcontentslineOrig=\addcontentsline
\def\addcontentsline#1#2#3{\bgroup
  \let\texttt=\textttOrig\addcontentslineOrig{#1}{#2}{#3}\egroup}
\let\markbothOrig\markboth
\def\markboth#1#2{\bgroup
  \let\texttt=\textttOrig\markbothOrig{#1}{#2}\egroup}
\let\markrightOrig\markright
\def\markright#1{\bgroup
  \let\texttt=\textttOrig\markrightOrig{#1}\egroup}

\usepackage{graphicx,grffile}
\makeatletter
\def\maxwidth{\ifdim\Gin@nat@width>\linewidth\linewidth\else\Gin@nat@width\fi}
\def\maxheight{\ifdim\Gin@nat@height>\textheight\textheight\else\Gin@nat@height\fi}
\makeatother
\setkeys{Gin}{width=\maxwidth,height=\maxheight,keepaspectratio}
\IfFileExists{parskip.sty}{%
\usepackage{parskip}
}{
\setlength{\parindent}{0pt}
\setlength{\parskip}{6pt plus 2pt minus 1pt}
}
\ifx\paragraph\undefined\else
\let\oldparagraph\paragraph
\renewcommand{\paragraph}[1]{\oldparagraph{#1}\mbox{}}
\fi
\ifx\subparagraph\undefined\else
\let\oldsubparagraph\subparagraph
\renewcommand{\subparagraph}[1]{\oldsubparagraph{#1}\mbox{}}
\fi

\title{radioactivedecay: A Python package for radioactive decay
calculations}

        \author[1]{Alex Malins}
          \author[2]{Thom Lemoine}

      \affil[1]{Center for Computational Science \& e-Systems (CCSE),
Japan Atomic Energy Agency (JAEA), 178-4-4 Wakashiba, Kashiwa, Chiba,
277-0871, Japan}
      \affil[2]{Whitman College, Walla Walla, Washington 99362, USA}
  \date{\vspace{-7ex}}

\begin{document}
\maketitle

\marginpar{

  \begin{flushleft}
  \sffamily\small

  {\bfseries DOI:} \href{https://doi.org/10.21105/joss.03318}{\color{linky}{10.21105/joss.03318}}

  \vspace{2mm}

  {\bfseries Software}
  \begin{itemize}
    \setlength\itemsep{0em}
    \item \href{https://github.com/openjournals/joss-reviews/issues/3318}{\color{linky}{Review}} \ExternalLink
    \item \href{https://github.com/radioactivedecay/radioactivedecay}{\color{linky}{Repository}} \ExternalLink
    \item \href{https://doi.org/10.5281/zenodo.6334651}{\color{linky}{Archive}} \ExternalLink
  \end{itemize}

  \vspace{2mm}

  \par\noindent\hrulefill\par

  \vspace{2mm}

  {\bfseries Editor:} \href{https://github.com/kellyrowland}{Kelly Rowland} \ExternalLink \\
  \vspace{1mm}
    {\bfseries Reviewers:}
  \begin{itemize}
  \setlength\itemsep{0em}
    \item \href{https://github.com/munkm}{@munkm}
    \item \href{https://github.com/shyamd}{@shyamd}
    \item \href{https://github.com/kellyrowland}{@kellyrowland}
    \end{itemize}
    \vspace{2mm}

  {\bfseries Submitted:} 30 April 2021\\
  {\bfseries Published:} 16 March 2022

  \vspace{2mm}
  {\bfseries License}\\
  Authors of papers retain copyright and release the work under a Creative Commons Attribution 4.0 International License (\href{http://creativecommons.org/licenses/by/4.0/}{\color{linky}{CC BY 4.0}}).

  \end{flushleft}
}

\hypertarget{summary}{%
\section{Summary}\label{summary}}

\texttt{radioactivedecay} is a Python package for radioactive decay
modelling. It contains functions to fetch decay data, define inventories
of nuclides and perform decay calculations. The default nuclear decay
dataset supplied with \texttt{radioactivedecay} is based on ICRP
Publication 107, which covers 1252 radioisotopes of 97 elements. The
code calculates an analytical solution to a matrix form of the decay
chain differential equations using double or higher precision numerical
operations. There are visualization functions for drawing decay chain
diagrams and plotting activity decay curves.

\hypertarget{statement-of-need}{%
\section{Statement of Need}\label{statement-of-need}}

Calculations for the decay of radioactivity and the ingrowth of progeny
underpin the use of radioisotopes in a wide range of research and
industrial fields, spanning from nuclear engineering, medical physics,
radiation protection, environmental science and archaeology to
non-destructive testing, mineral prospecting, food preservation,
homeland security and defence. \texttt{radioactivedecay} is an open
source, cross-platform package for decay calculations and visualization.
It supports decay chains with branching decays and metastable nuclear
isomers. It includes a high numerical precision decay calculation mode,
which resolves numerical problems with using double-precision
floating-point numbers to calculate decay chains involving radionuclides
with disparate half-lives (Bakin et al., 2018).

This set of features distinguishes \texttt{radioactivedecay} from other
commonly-used decay packages, such as \texttt{Radiological\ Toolbox}
(Hertel et al., 2015) and \texttt{PyNE} (Scopatz et al., 2012).
\texttt{Radiological\ Toolbox} is a closed-source Windows application,
so it is not easily scriptable and its use of double-precision
arithmetic makes it susceptible to numerical round-off errors.
\texttt{PyNE} uses approximations to help mitigate numerical issues,
however these may potentially affect accuracy. Moreover as of
\texttt{v0.7.5}, \texttt{PyNE} does not correctly model metastable
nuclear isomers within decay chains, which means, for example, it cannot
simulate the production of \(^{99m}\textrm{Tc}\) from
\(^{99}\textrm{Mo}\) for medical imaging applications.

\hypertarget{theory-and-implementation}{%
\section{Theory and Implementation}\label{theory-and-implementation}}

\texttt{radioactivedecay} implements the solution to the decay
differential equations outlined by Amaku et al. (2010). If vector
\(\mathbf{N}\) contains the number of atoms of each radionuclide in a
system, its elements \(N_i\) can be ordered such that no progeny (either
first or subsequent generation) of radionuclide \(i\) has itself an
index lower than \(i\). Ordering in this manner is possible because
natural radioactive decay processes do not increase the mass number of
the decaying radionuclide, and there are no cyclic decay chains where
radionuclide \(i\) can decay to other radionuclides then reform itself
(Ladshaw et al., 2020). Note metastable nuclear isomers have distinct
indices from their ground states in \(\mathbf{N}\).

The radioactive decay chain differential equations expressed in matrix
form are:

\begin{equation}
\frac{\mathrm{d}\mathbf{N}}{\mathrm{d}t} = \varLambda \mathbf{N}.
\label{eq:diff_eq}
\end{equation}

\(\varLambda\) is a lower triangular matrix with elements:

\begin{equation}
\varLambda_{ij} =
\begin{cases}
0 & \textrm{for }  i < j,\\
-\lambda_{j} & \textrm{for }  i = j,\\
b_{ji}\lambda_{j} & \textrm{for }  i > j.
\end{cases}
\end{equation}

\(\lambda_{j}\) is the decay constant of radionuclide \(j\), and
\(b_{ji}\) is the branching fraction from radionuclide \(j\) to \(i\).
\(\varLambda\) is diagonalizable so its eigendecomposition can be used to
rewrite \autoref{eq:diff_eq} as:

\begin{equation}
\frac{\mathrm{d}\mathbf{N}}{\mathrm{d}t} = C \varLambda_d C^{-1} \mathbf{N}.
\label{eq:diff_eq_rewrite}
\end{equation}

\(\varLambda_d\) is a diagonal matrix whose elements are the negative decay
constants, i.e.~\(\varLambda_{dii} = -\lambda_{i}\). Matrix \(C\) and its
inverse \(C^{-1}\) are both lower triangular matrices that are
calculated as:

\begin{equation}
C_{ij} =
\begin{cases}
0 & \text{for }  i < j,\\
1 & \text{for }  i = j,\\
\frac{\sum_{k=j}^{i-1}\varLambda_{ik}C_{kj}}{\varLambda_{jj} - \varLambda_{ii}} & \text{for }  i > j,
\end{cases}
\quad\text{and}\quad
C^{-1}_{ij} =
\begin{cases}
0 & \text{for }  i < j,\\
1 & \text{for }  i = j,\\
-\sum_{k=j}^{i-1} C_{ik} C^{-1}_{kj} & \text{for }  i > j.
\end{cases}
\label{eq:c}
\end{equation}

The analytical solution to \autoref{eq:diff_eq_rewrite} given an initial
condition of \(\mathbf{N}(0)\) at \(t=0\) is:

\begin{equation}
\mathbf{N}(t) = C e^{\varLambda_{d} t} C^{-1} \mathbf{N}(0).
\label{eq:solution}
\end{equation}

\(e^{\varLambda_{d} t}\) is a diagonal matrix with elements
\(e^{\varLambda_{d} t}_{ii} = e^{-\lambda_i t}\). \texttt{radioactivedecay}
evaluates \autoref{eq:solution} upon each call for a decay calculation.

Matrices \(C\) and \(C^{-1}\) are independent of time so they are
pre-calculated and imported from files into \texttt{radioactivedecay}.
\(C\) and \(C^{-1}\) are stored in sparse matrix data structures to
minimize memory use and maximize efficiency when computing the matrix
multiplications in \autoref{eq:solution}. For decay calculations with
double-precision floating-point operations, \(C\) and \(C^{-1}\) are
stored in \texttt{SciPy} (Virtanen et al., 2020) Compressed Sparse Row
(CSR) matrix data structures. Conversely, they are stored in
\texttt{SymPy} (Meurer et al., 2017) SparseMatrix data structures for
high numerical precision calculations.

The high numerical precision decay calculation mode resolves numerical
issues arising from using double-precision floating-point numbers for
decay calculations for chains containing nuclides with disparate
half-lives. One example is the decay chain for \(^{254}\textrm{Es}\),
which contains \(^{238}\textrm{U}\) (4.468 billion year half-life) and
\(^{214}\textrm{Po}\) (\(t_{1/2}\) is 164.3 \(\mu\)s half-life). This a
20 orders of magnitude difference in half-life. Loss of numerical
precision inevitably occurs when evaluating the off-diagonal elements of
\(C\) and \(C^{-1}\) in \autoref{eq:c} with double-precision
floating-point numbers (which hold approximately 15 decimal places of
numerical precision). Note loss of precision also occurs in the converse
scenario, i.e.~when a decay chain contains radionuclides with similar
half-lives. However this scenario does not occur in the ICRP Publication
107 decay dataset, as the relative difference between half-lives of any
two radionuclides in the same decay chain is always greater than 0.1\%.

The default operation of the high precision decay mode is to evaluate
\autoref{eq:solution} using floating-point numbers with 320 significant
figures of precision. This is sufficient precision to ensure accurate
results for any physically relevant decay calculation users may wish to
perform. Moreover, computations in the high precision mode are still
fast, taking less than one second on a notebook equipped with an Intel
Core i5-8250U processor.

\hypertarget{decay-atomic-mass-datasets}{%
\section{Decay \& Atomic Mass
Datasets}\label{decay-atomic-mass-datasets}}

The default dataset supplied with \texttt{radioactivedecay} uses decay
data from ICRP Publication 107 (Eckerman \& Endo, 2008) and atomic
masses from the Atomic Mass Data Center (AMDC) (Huang et al., 2021;
Kondev et al., 2021; Wang et al., 2021). Endo et al. (2005) and Endo \&
Eckerman (2007) describe the development of the ICRP Publication 107
decay dataset. Raw data from ICRP 107 and AMDC were converted into
dataset files suitable for \texttt{radioactivedecay} in a Jupyter
\href{https://github.com/radioactivedecay/datasets}{notebook}. Along
with \texttt{SciPy} and \texttt{SymPy} versions of the sparse matrices
\(C\) and \(C^{-1}\), the dataset files contain radionuclide half-lives,
decay constants, progeny, branching fractions, decay modes and atomic
masses. Although there is a default dataset, \texttt{radioactivedecay}
allows the import and use other decay data.

\hypertarget{main-functionality}{%
\section{Main Functionality}\label{main-functionality}}

\begin{figure}
\centering
\includegraphics{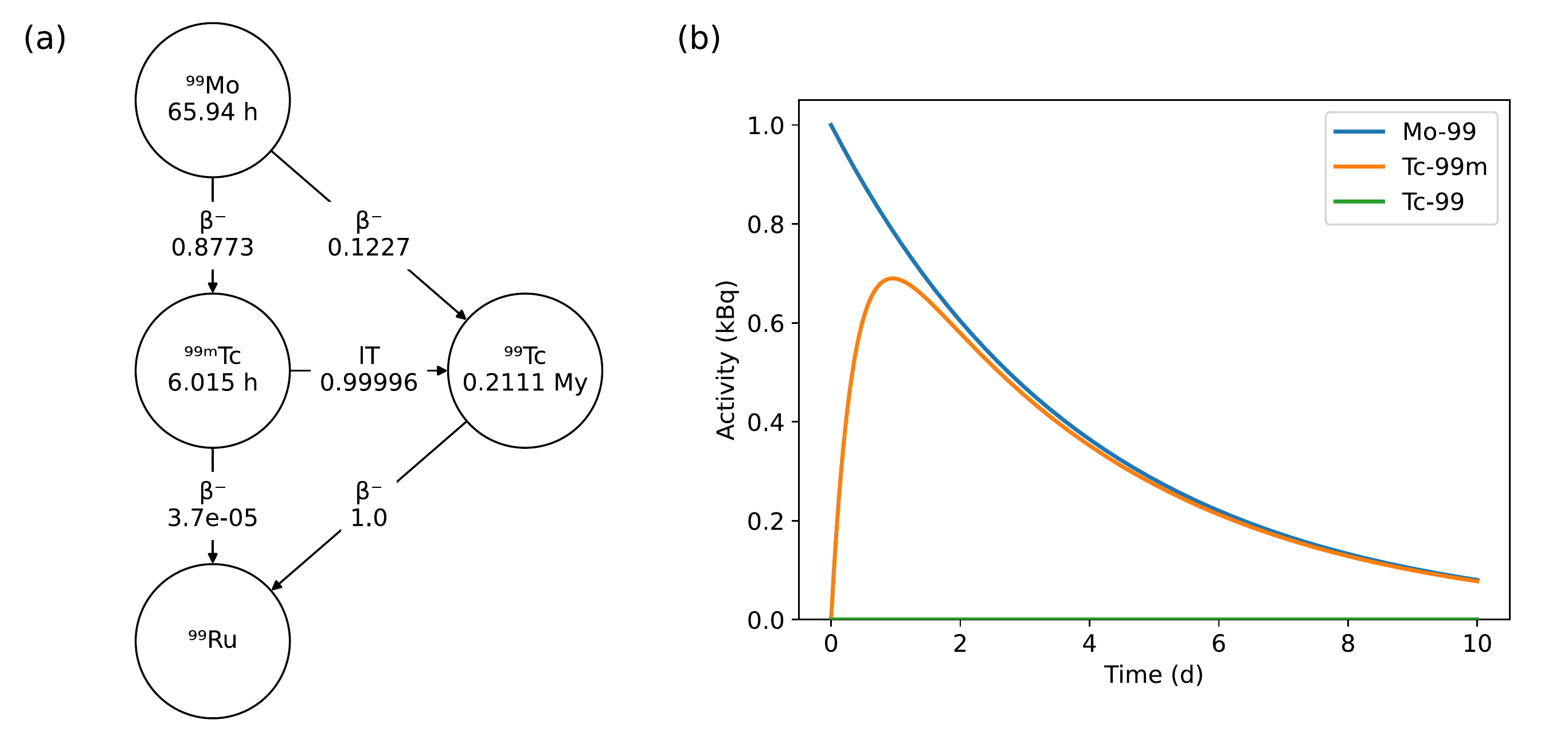}
\caption{Examples of the plotting capabilities of
\texttt{radioactivedecay}: (a) Decay chain diagram for molybdenum-99.
(b) Graph showing the decay of 1 kBq of \(^{99}\textrm{Mo}\) along with
the ingrowth of \(^{99m}\textrm{Tc}\) and a trace quantity of
\(^{99}\textrm{Tc}\).\label{fig:decay_diags}}
\end{figure}

The main functionality of \texttt{radioactivedecay} is based around
\texttt{Nuclide}, \texttt{Inventory} and \texttt{InventoryHP} classes.
The \texttt{Nuclide} class is used for fetching atomic and decay data
about a single nuclide, such as its atomic mass, half-life, decay modes,
progeny and branching fractions. It creates diagrams of the nuclide's
decay chain (ex. \autoref{fig:decay_diags}(a)) using the
\texttt{NetworkX} library (Hagberg et al., 2008).

An \texttt{Inventory} can contain multiple nuclides, each with an
associated quantity (the number of atoms of the nuclide). Nuclides can
be stable or radioactive. The \texttt{decay()} method calculates the
decay of the radioactive nuclides in an \texttt{Inventory}, adding any
ingrown progeny automatically. The \texttt{numbers()},
\texttt{activities()}, \texttt{masses()}, and \texttt{moles()} methods
output the inventory of nuclides as different quantities using the
atomic data stored in the decay dataset. Additional
\texttt{activity\_fractions()}, \texttt{mass\_fractions()}, and
\texttt{mole\_fractions()} methods provide the relative amounts of each
nuclide in the inventory with respect to different quantities. Plots can
be made of the variation of nuclide activities, masses and moles over
time (ex. \autoref{fig:decay_diags}(b)) using \texttt{Matplotlib}
(Hunter, 2007).

The \texttt{InventoryHP} class is the high numerical precision
complement of the \texttt{Inventory} class. It has the same API as the
\texttt{Inventory} class, but uses \texttt{SymPy} high numerical
precision routines for all calculations.

\hypertarget{validation}{%
\section{Validation}\label{validation}}

Decay calculations with \texttt{radioactivedecay\ v0.4.2} were
cross-checked against \texttt{Radiological\ Toolbox\ v3.0.0} (Hertel et
al., 2015) and \texttt{PyNE\ v0.7.5} (Scopatz et al., 2012) (see Jupyter
notebooks in the
\href{https://github.com/radioactivedecay/comparisons}{comparisons
repository}). \texttt{Radiological\ Toolbox} employs the ICRP
Publication 107 decay data. Fifty radionuclides were randomly selected
and a decay calculation was performed for 1 Bq of each for a random
decay time within a factor of \(10^{-3}\) to \(10^{3}\) of the
half-life. Differences between decayed activities reported by each code
were within 1\% of each other in 64\% of cases. Discrepancies greater
than 1\% were attributed to rounding differences, erroneous results from
\texttt{Radiological\ Toolbox}, or numerical issues relating to decay
chains containing radionuclides with disparate half-lives.

A dataset was prepared for \texttt{radioactivedecay} with the same
Evaluated Nuclear Structure Data File (ENSDF, 2019) decay data as used
by \texttt{PyNE\ v0.7.5}. Bugs in \texttt{PyNE\ v0.7.5} cause incorrect
decay calculation results for chains containing metastable nuclear
isomers, \(^{183}\textrm{Pt}\), \(^{172}\textrm{Ir}\) or
\(^{152}\textrm{Lu}\). Thus the affected chains were not used for the
comparisons. The decay of 1 Bq of every radionuclide was calculated for
multiple decay times varying from \(0\) to \(10^{6}\) times the
radionuclide's half-life. The absolute difference between the decayed
activities reported by each code was less than \(10^{-13}\) Bq. Relative
differences depended on the magnitude of the activity. Relative errors
of greater than 0.1\% only occurred when the calculated activity was
less than \(2.5\times10^{-11}\) Bq, i.e.~10 orders of magnitude smaller
than the initial activity of the parent radionuclide. The discrepancies
between the two codes were attributed to methodological differences for
computing decay chains with radionuclides with large disparities between
half-lives, and numerical issues arising from double-precision
floating-point operations.

\hypertarget{limitations}{%
\section{Limitations}\label{limitations}}

\texttt{radioactivedecay} does not model neutronics, so cannot evaluate
radioactivity produced from activations or induced fission. It does not
support external sources of radioactivity input or removal from an
inventory over time. Caution is required if decaying backwards in time,
as this can cause floating-point overflows when computing the
exponential terms in \autoref{eq:solution}.

There are also some limitations associated with the ICRP Publication 107
decay dataset. It does not contain data for the radioactivity produced
from spontaneous fission decay pathways and the minor decay pathways of
some radionuclides. More details on limitations are available in the
\href{https://radioactivedecay.github.io/overview.html\#limitations}{documentation},
Endo et al. (2005), and Endo \& Eckerman (2007).

\hypertarget{acknowledgements}{%
\section{Acknowledgements}\label{acknowledgements}}

We thank Mitsuhiro Itakura, Kazuyuki Sakuma, colleagues in JAEA's Center
for Computational Science \& Systems, Wolfgang Kerzendorf \& Bernardo
Gameiro for their support for this project. We thank Kenny McKee, Daniel
Jewell, Ezequiel Passaro, Hunter Ratliff \& Jayson Vavrek for helpful
suggestions, and Bjorn Dahlgren, Anthony Scopatz \& Jonathan Morrell for
their work on radioactive decay calculation software. We also thank the
editors and reviewers at the Journal of Open Source software for
constructive comments.

\hypertarget{references}{%
\section*{References}\label{references}}
\addcontentsline{toc}{section}{References}

\hypertarget{refs}{}
\begin{CSLReferences}{1}{0}
\leavevmode\hypertarget{ref-Amaku2010}{}%
Amaku, M., Pascholati, P. R., \& Vanin, V. R. (2010). {Decay chain
differential equations: Solution through matrix algebra}. \emph{Computer
Physics Communications}, \emph{181}(1), 21--23.
\url{https://doi.org/10.1088/0952-4746/26/3/N02}

\leavevmode\hypertarget{ref-Bakin2018}{}%
Bakin, R. I., Kiselev, A. A., Shvedov, A. M., \& Shikin, A. V. (2018).
{Computational Errors in the Calculation of Long Radioactive Decay
Chains}. \emph{Atomic Energy}, \emph{123}(6), 406--411.
\url{https://doi.org/10.1007/s10512-018-0360-2}

\leavevmode\hypertarget{ref-ICRP107}{}%
Eckerman, K. F., \& Endo, A. (2008). {ICRP 107: Nuclear Decay Data for
Dosimetric Calculations}. \emph{Annals of the ICRP}, \emph{38}(3), 119.
\url{https://doi.org/10.1016/j.icrp.2008.10.003}

\leavevmode\hypertarget{ref-Endo2007}{}%
Endo, A., \& Eckerman, K. F. (2007). \emph{{JAEA-Data/Code 2007-021:
Nuclear Decay Data for Dosimetry Calculation - Data for Radionuclides
with Half-lives Less than 10 Minutes}}.
\url{https://doi.org/10.11484/jaea-data-code-2007-021}

\leavevmode\hypertarget{ref-Endo2005}{}%
Endo, A., Yamaguchi, Y., \& Eckerman, K. F. (2005). \emph{{JAERI 1347:
Nuclear Decay Data for Dosimetry Calculation; Revised data of ICRP
Publication 38}}. \url{https://doi.org/10.11484/jaeri-1347}

\leavevmode\hypertarget{ref-ENSDF}{}%
ENSDF. (2019). \emph{{From ENSDF database as of October 4, 2019. Version
available at:}} \url{https://www.nndc.bnl.gov/ensarchivals/}

\leavevmode\hypertarget{ref-Hagberg2008}{}%
Hagberg, A. A., Schult, D. A., \& Swart, P. J. (2008). {Exploring
network structure, dynamics, and function using NetworkX}.
\emph{Proceedings of the 7th Python in Science Conference (SciPy2008)},
11--15. \url{http://conference.scipy.org/proceedings/SciPy2008/paper_2/}

\leavevmode\hypertarget{ref-Hertel2015}{}%
Hertel, N. E., Eckerman, K. F., \& Sun, C. (2015). {Radiological Tookbox
3.0.0}. \emph{Transactions of the American Nuclear Society}, \emph{113},
977--980. \url{https://www.ans.org/pubs/transactions/article-38022/}

\leavevmode\hypertarget{ref-Huang2021}{}%
Huang, W. J., Wang, M., Kondev, F. G., Audi, G., \& Naimi, S. (2021).
The {AME} 2020 atomic mass evaluation (i). Evaluation of input data, and
adjustment procedures. \emph{Chinese Physics C}, \emph{45}(3), 030002.
\url{https://doi.org/10.1088/1674-1137/abddb0}

\leavevmode\hypertarget{ref-Hunter2007}{}%
Hunter, J. D. (2007). {Matplotlib: A 2D graphics environment}.
\emph{Computing in Science \& Engineering}, \emph{9}(3), 90--95.
\url{https://doi.org/10.1109/MCSE.2007.55}

\leavevmode\hypertarget{ref-Kondev2021}{}%
Kondev, F. G., Wang, M., Huang, W. J., Naimi, S., \& Audi, G. (2021).
The {NUBASE}2020 evaluation of nuclear physics properties. \emph{Chinese
Physics C}, \emph{45}(3), 030001.
\url{https://doi.org/10.1088/1674-1137/abddae}

\leavevmode\hypertarget{ref-Ladshaw2020}{}%
Ladshaw, A., Wiechert, A. I., Kim, Y., Tsouris, C., \& Yiacoumi, S.
(2020). {Algorithms and algebraic solutions of decay chain differential
equations for stable and unstable nuclide fractionation}. \emph{Computer
Physics Communications}, \emph{246}, 106907.
\url{https://doi.org/10.1016/j.cpc.2019.106907}

\leavevmode\hypertarget{ref-Meurer2017}{}%
Meurer, A., Smith, C. P., Paprocki, M., ?ertik, O., Kirpichev, S. B.,
Rocklin, M., Kumar, A., Ivanov, S., Moore, J. K., Singh, S., Rathnayake,
T., Vig, S., Granger, B. E., Muller, R. P., Bonazzi, F., Gupta, H.,
Vats, S., Johansson, F., Pedregosa, F., \ldots{} Scopatz, A. (2017).
{SymPy: symbolic computing in Python}. \emph{PeerJ Computer Science},
\emph{3}, e103. \url{https://doi.org/10.7717/peerj-cs.103}

\leavevmode\hypertarget{ref-Scopatz2012}{}%
Scopatz, A. M., Romano, P. K., Wilson, P. P. H., \& Huff, K. D. (2012).
{PyNE: Python for nuclear engineering}. \emph{Transactions of the
American Nuclear Society}, \emph{107}, 985--987.
\url{https://www.ans.org/pubs/transactions/article-14978}

\leavevmode\hypertarget{ref-Virtanen2020}{}%
Virtanen, P., Gommers, R., Oliphant, T. E., Haberland, M., Reddy, T.,
Cournapeau, D., Burovski, E., Peterson, P., Weckesser, W., Bright, J.,
Walt, S. J. van der, Brett, M., Wilson, J., Millman, K. J., Mayorov, N.,
Nelson, A. R. J., Jones, E., Kern, R., Larson, E., \ldots{}
Vazquez-Baeza, Y. (2020). {SciPy 1.0: fundamental algorithms for
scientific computing in Python}. \emph{Nature Methods}, \emph{17}(3),
261--272. \url{https://doi.org/10.1038/s41592-019-0686-2}

\leavevmode\hypertarget{ref-Wang2021}{}%
Wang, M., Huang, W. J., Kondev, F. G., Audi, G., \& Naimi, S. (2021).
The {AME} 2020 atomic mass evaluation ({II}). Tables, graphs and
references. \emph{Chinese Physics C}, \emph{45}(3), 030003.
\url{https://doi.org/10.1088/1674-1137/abddaf}

\end{CSLReferences}

\end{document}